\documentclass[a4paper,reprint,notitlepage,superscriptaddress,aps,pra]{revtex4-1}

\usepackage{amsmath,amssymb,amsfonts}

\usepackage{graphicx}
\usepackage[colorlinks,
	linkcolor=red,
	citecolor=blue,
	urlcolor=red]{hyperref}

\usepackage{cleveref}

\renewcommand{\t}[1]{\mathrm{#1}}
\newcommand{\abs}[1]{\vert {#1} \vert}
\renewcommand{\arg}[1]{\mathrm{arg}\,}

\begin{document}

\title{Sub-Hertz Optomechanically-Induced Transparency}

\author{T. Bodiya}
\affiliation{LIGO Laboratory, Massachusetts Institute of Technology, Cambridge, MA 02139}
\affiliation{Laser Applications Group, Lincoln Laboratory, Lexington, MA 02474}
\author{V. Sudhir}
\email{vivishek.sudhir@ligo.org}
\affiliation{LIGO Laboratory, Massachusetts Institute of Technology, Cambridge, MA 02139}
\author{C. Wipf}
\affiliation{LIGO Laboratory, Massachusetts Institute of Technology, Cambridge, MA 02139}
\affiliation{LIGO Laboratory, California Institute of Technology, Pasadena, CA 91125}
\author{N. Smith}
\author{A. Buikema}
\affiliation{LIGO Laboratory, Massachusetts Institute of Technology, Cambridge, MA 02139}
\author{A. Kontos}
\affiliation{LIGO Laboratory, Massachusetts Institute of Technology, Cambridge, MA 02139}
\affiliation{Bard College, Annandale-on-Hudson, NY 12504}
\author{H. Yu}
\author{N. Mavalvala}
\email{nergis@ligo.mit.edu}
\affiliation{LIGO Laboratory, Massachusetts Institute of Technology, Cambridge, MA 02139}

\begin{abstract}
Optical interferometers with suspended mirrors are the archetype of all current audio-frequency gravitational-wave
detectors. The radiation pressure interaction between the motion of the mirror and the circulating optical
field in such interferometers represents a pristine form of light-matter coupling, largely due to 30 years of effort
in developing high quality optical materials with low mechanical dissipation.
However, in all current suspended interferometers, the radiation pressure interaction is too weak to be useful
as a resource, and too strong to be neglected as a noise source.
Here, we demonstrate a meter-long interferometer with suspended mirrors of effective mass $~ 125\, \t{g}$,
where the radiation pressure interaction is enhanced by strong optical pumping to realize a cooperativity of $50$.
We probe this regime by observing optomechanically-induced transparency of a weak on-resonant probe.
The low resonant frequency and high-Q of the mechanical oscillator allows us to demonstrate transparency windows 
barely $100\, \t{mHz}$ wide at room temperature.
Together with a near-unity ($\sim 99.9\%$) out-coupling efficiency, our system saturates the theoretical
delay-bandwidth product, rendering it an optical buffer capable of seconds-long storage times.
\end{abstract}

\maketitle

Interferometers with suspended end-mirror cavities are one of today's most sensitive instruments \cite{ligo_det}.
Suspending the optics
isolate them from technical noises of seismic and anthropic origin. Once classical noises are mitigated,
the sensitivity of the interferometer increases with the intensity of the optical field circulating within.
However, high-power operation 
is limited by classical and quantum mechanical effects of radiation pressure \cite{Brag67,Brag68,Cav80}.
The longitudinal motion of the suspended end-mirror, considered as a harmonic oscillator, is susceptible
to two effects arising from the coupling between its motion and the circulating optical field.
Classically, a radiation pressure force that depends on the oscillator position 
-- due to feedback from the optical (cavity) delay -- 
can lead to parametric instability \cite{CorMav06,Evans15}.
Quantum mechanically, the fluctuations in the number of photons recoiling off of the end-mirror --
quantum radiation pressure noise -- can perturb the oscillator \cite{Purd13,Wils15,Teuf16,CriCor18}.
Whilst the former effect can be described as a modification of the susceptibility of the oscillating
mirror due to its coupling to the optical field, the latter is a fluctuating force originating from the same coupling.
Generally it can be shown that the two effects scale identically with power \cite{Clerk10}.
This scaling is described by the dimensionless radiation pressure coupling strength, quantified by the
\emph{cooperativity} $C$ (to be defined below).
At present in Advanced LIGO, the radiation pressure coupling between a higher order mechanical mode 
and transverse optical mode has been shown to be strong enough ($C \approx 1$) to initiate parametric 
instability \cite{Evans15},
yet weak enough that quantum radiation pressure noise is buried tantalizingly beneath technical 
noises \cite{ligo_det,ligo_corr}.

Here we demonstrate a suspended-mirror interferometer with a radiation pressure cooperativity an order of magnitude
larger ($C\approx 50$) than what has previously been directly observed in such an 
instrument \cite{CorMav06,Miya06,Corb07a,Corb07b,Evans15}.
In contrast to nano-fabricated optomechanical systems \cite{AspMey12,AspKipMar14},
our system consists of a mechanical oscillator of effective mass $125\, \t{g}$
-- 9 orders of magnitude more massive -- 
susceptible to the recoil-type radiation pressure coupling,
and an interferometer that is formed by a 1m long suspended cavity,
both of which make it a mock-up of an Advanced LIGO arm.
This system serves as a general experimental platform for audio-band optomechanics
in a radiation-pressure-dominated regime \cite{CorMav06,Corb07a,Corb07b}.

\begin{figure}[t!]
\includegraphics[width=0.99\columnwidth]{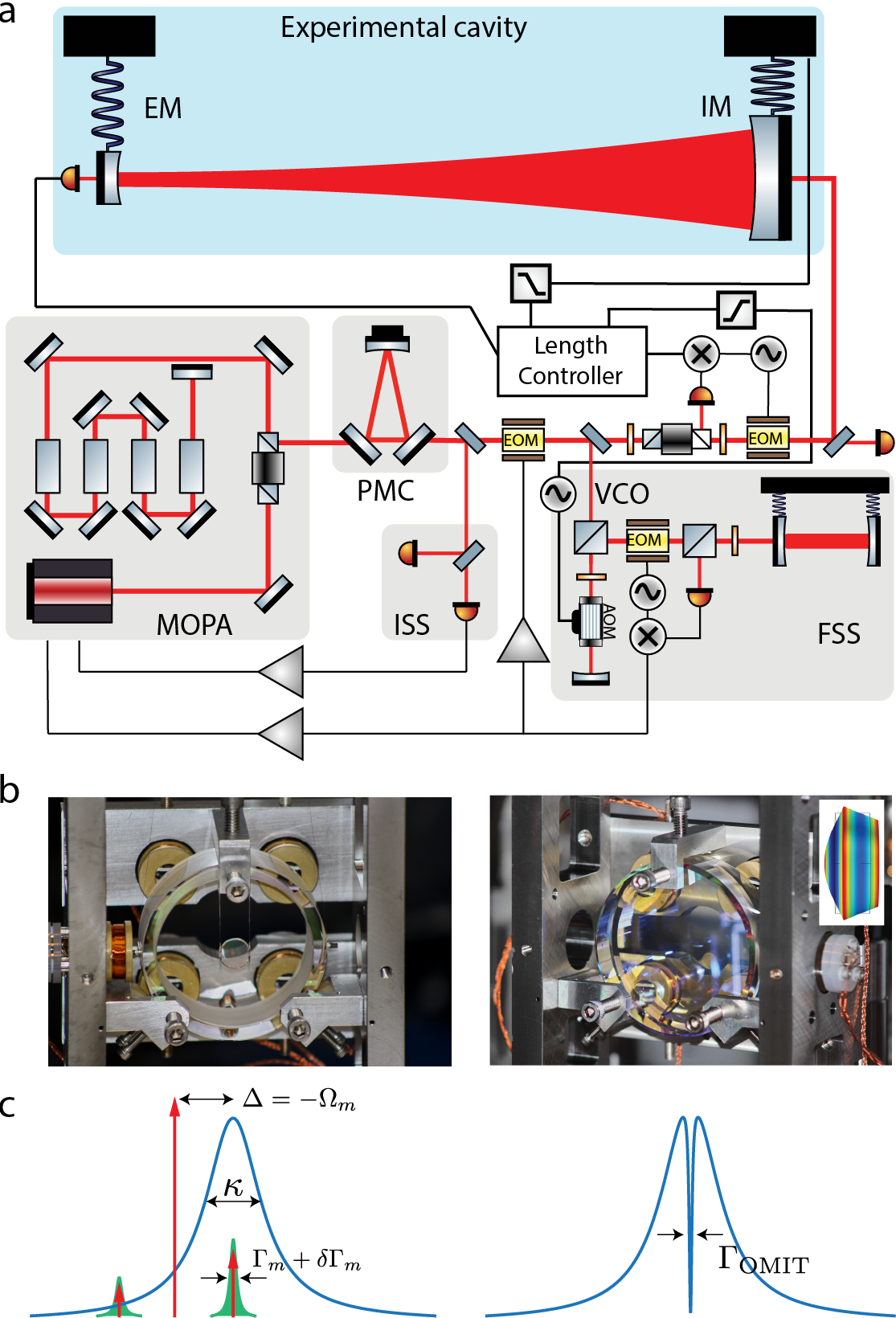}
\caption{\label{fig:expt_overview}
  Experimental system and schematic.
  (a) Laser light for the experiment is derived from a master-oscillator power amplifier (MOPA), that was
  developed for Initial LIGO.
  The 10 W output, at 1064 nm, is spatially filtered by a pre-mode cleaner (PMC);
  subsequently, its intensity (ISS) and frequency (FSS) are stabilized using respective servos.
  The light then enters the experimental cavity, whose reflection and transmission are monitored and used
  for stabilizing the laser-cavity detuning.
  (b) Left: Picture of the end mirror (EM), consisting of a $1.27$ cm diameter mirror, weighing $1$ gm, suspended
  as a double pendulum.
  Right: Picture of the input mirror (IM), consisting of a $7.3$ cm diameter mirror, weighing $250$ g, suspended
  as a single pendulum. Inset shows finite element model of the mirror's fundamental drumhead mode, which is the
  mechanical oscillator relevant to this work.
  (c) Simple scheme of optomechanically-induced transparency (OMIT). Left shows the bare cavity response (blue), 
  pumped by laser light red-detuned by a mechanical linewidth ($\Delta = -\Omega_m$), and its two sidebands filtered
  by the cavity; the radiation-pressure-induced displacement is transduced as additional sidebands (green), which
  interferes with the injected sidebands. Right shows the resulting effective cavity response.  
  }
\end{figure}

The basic challenge of operating such an interferometer in the high cooperativity regime is the ability to
store enough photons in the cavity to amplify the radiation pressure force without destabilizing it by other means.
In order to retain sufficient photons in the cavity with practical input laser powers, it is
necessary that the end mirrors' optical losses -- already at a state-of-the art level of a few parts per
million (ppm) --
be diluted by elongating the cavity.
Operating a long cavity with suspended mirrors introduces additional challenges that must be overcome.
In particular, maintaining alignment requires seismic isolation whose fundamental suspension mode is
low frequency ($\sim 1\, \t{Hz}$). Thermal noise requirements demand the suspending fibers be thin.
Thus, the suspensions are necessarily ``soft'', and their torsional modes
are susceptible to a radiation-pressure torque instability 
\cite{SoliBar91,SidSig06} -- an effect that
has been a limiting factor in the high power operation of suspended interferometers \cite{HirSaul10,SakMiy10,DooBar13}.
On the one hand, the magnitude of the resonant round-trip gain for the radiation-pressure-induced longitudinal 
coupling between cavity frequency and length is the cooperativity, 
\begin{equation}\label{eq:C}
  C \equiv \frac{4g^2}{\kappa \Gamma_m},
\end{equation}
where $g = (\partial \omega_c/\partial x) \sqrt{\hbar n_c/(2m\Omega_m)}$, is the optomechanical coupling rate
for an end-mirror of effective mass $m$, oscillating at frequency $\Omega_m$, which leads to a cavity frequency
fluctuation $\partial \omega_c/\partial x = \omega_c/L$ for a cavity of length $L$ and decay rate $\kappa$, 
loaded with $n_c$ photons on average. 
On the other hand, the gain for the torque coupling scales as $\kappa n_c L/M$, where $M$
is the moment of inertia \cite{SidSig06}.
The ratio of the round-trip gains of the
two processes (longitudinal and torsional radiation pressure) scale as $m^{-2}L^{-3}$ -- heavier mirrors reduce 
the susceptibility to radiation pressure force, while
longer cavities enhance the effect of radiation pressure torque.

The competing demands of being radiation pressure dominated longitudinally while still maintaining angular stability 
are met in our experimental system, depicted in \Cref{fig:expt_overview}. The optical cavity of interest is formed by
two mirrors -- an end mirror (EM) weighing $1\, \t{g}$ with a transmission of 3 ppm, and an
input mirror (IM) weighing $250\, \t{g}$ with a transmission 800 ppm -- suspended on $1\, \t{Hz}$ pendulums
placed $1\, \t{m}$ apart (this gives a cavity of linewidth $\kappa \approx 2\pi \cdot 21\, \t{kHz}$). 
In order to suppress extraneous beam-pointing noise, the experiment is mounted on an actively damped seismic
vibration isolation platform, similar to the one used in Advanced LIGO \cite{SEI15}, which attenuates ground
motion to a level of $\sim 10^{-9}\, \t{m/\sqrt{Hz}}$ above a few Hz.
The cavity is driven by laser light from a master-oscillator power amplifier (MOPA) capable of delivering
10 W of continuous output at 1064 nm \cite{PSL99}.
Extraneous power fluctuations are reduced to a relative intensity noise of $\sim 10^{-8}/\t{\sqrt{Hz}}$ above 
100 Hz; this, together with carefully centering of the beam position on the mirrors (by minimizing the 
transduction of the suspension pitch mode onto the phase of the light leaking out of the cavity), reduces 
torque fluctuations to insignificant levels.
The spatial mode of the laser is cleaned by passing it through a pre-mode cleaner, so as to prevent extraneous signals
from higher order modes in the quadrant photodetectors (not shown in figure) that are used to feedback-stabilize
cavity alignment.
The laser is frequency-stabilized to an independent reference cavity.
In this configuration, optical torque instabilities
are not expected to limit high power operation, and it should be possible to realize $C \gtrsim 100$.

\begin{figure}[t!]
\includegraphics[width=\columnwidth]{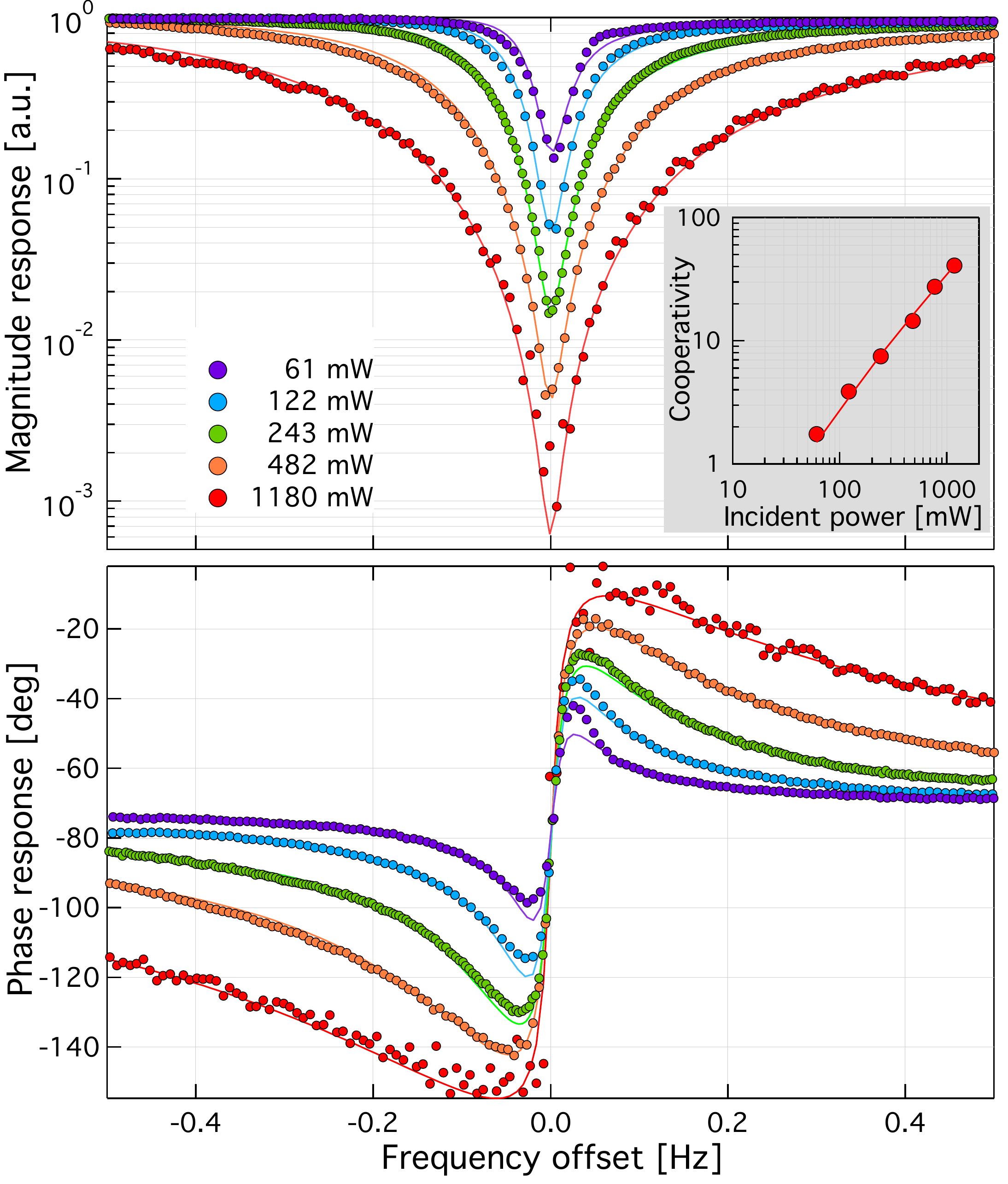}
\caption{\label{fig:omit}
  Sub-hertz optomechanically-induced transparency: main figure shows zoom-in of the magnitude (top)
  and phase response (bottom) of the cavity transmission in a 1 Hz span around resonance at several values
  of the incident power (legend).
  Solid lines show model curves.
  Inset shows the optomechanical cooperativity inferred from the model.
  }
\end{figure}

Exploiting the slew of technical capabilities originally developed for LIGO, we observe 
optomechanically-induced transparency (OMIT) \cite{AgHu10,omit10,omit11,Teuf11,XioWu18} due to the coupling
between the drumhead mode of the IM (see \cref{fig:expt_overview}b) and the intracavity field. 
Modulating the incident laser at a variable frequency offset $\Omega$ from its carrier  creates an intracavity
radiation pressure at the same frequency. When $\Omega=\Omega_m \approx 2\pi\cdot 27\, \t{kHz}$,
the drumhead mode is resonantly excited by this intracavity radiation pressure force; this
displacement gets transduced as phase-modulation sidebands back on the intracavity field, which can interfere with
the injected modulation sideband. When the carrier is detuned from the cavity resonance by the mechanical frequency
(i.e. $\Delta =\Omega_m$), the interference is perfectly destructive. 
In the experiment we probe the cavity response from within the feedback loop used to stabilize its length. 
First, we acquire lock of the cavity length to the laser by using a Pound-Drever-Hall signal in reflection 
(see \cref{fig:expt_overview}a); we then hand over the length control from the reflection signal to the 
transmission signal in order to be able to red-detune by $\Delta =-\Omega_m$.
The cavity response is then probed by frequency-modulation sidebands generated by dithering the 
frequency servo error point, and demodulating the signal out of the photodetector in transmission. 
We verify the detuning by fits to the broadband response of the cavity (see \cite{supp_info}).
Finally, we measure the response in the vicinity of the cavity resonance in a high-resolution scan to observe
the narrow OMIT feature. However, the response observed in this manner needs to be corrected
for the response of the frequency stabilization loop through which both the probe and the measurement
are made (see \cite{supp_info} for details).
\Cref{fig:omit} shows the magnitude and phase of this corrected response as the power in the incident laser
is increased.
The gross features of the cavity transmission can be understood from the approximate model \cite{supp_info},
\begin{equation}\label{eq:T}
  T[\Omega] \approx 
    T_0[\Omega]
    \left(1-\frac{\delta \Gamma_m/2}{(\Gamma_m+\delta \Gamma_m)/2 +i(\Omega_m -\Omega)} \right),
\end{equation}
where, $T_0[\Omega]$, is the cavity transmission without any optomechanical coupling, and 
$\delta \Gamma_m \approx \Gamma_m\, C_\t{eff}$, is the optically-damped contribution to the mechanical decay rate;
here $C_\t{eff}=C/[1+(\kappa/4\Omega_m)^2]$ is the effective cooperativity taking into account the finite
sideband resolution $\Omega_m/\kappa \approx 1.4$, which gives, $C_\t{eff}\approx 0.97\cdot C$.
The bare cavity transmission $T_0$ is a Lorentzian of width $\kappa$, while the second factor
describes an OMIT window of width given by the effective mechanical linewidth, on cavity resonance.
As the cooperativity increases, $T[\Omega_m]/T_0[\Omega_m] \approx 1/(1+C_\t{eff}) \rightarrow 0$.
The expression in \cref{eq:T} is however an approximation that disregards the finite sideband resolution; 
in fact, it only describes the contribution to the transmitted photodetector signal that arises from the upper-sideband
of the intracavity field.
A full model accounting for both sidebands is shown as the fits in \cref{fig:omit} (see \cite{supp_info} for
the full model). 
These fits allow us to extract the cooperativity, shown as the inset in the figure. 
At the highest incident power of 1.2 W, we realize $C\approx 50$.

Our results of $\Gamma_\t{OMIT}/2\pi \lesssim 100$ mHz demonstrate the narrowest OMIT windows yet 
observed (a recent experiment at dilution refrigerator temperature is comparable \cite{YuStee15}).
This is largely due to the ability to operate an optomechanical system in the high-cooperativity regime
using a mechanical oscillator with a long decay time. We achieve this by using a low frequency oscillator
featuring a high intrinsic mechanical quality factor
of $Q_m = \Omega_m/\Gamma_m \approx 10^6$, consistent with expectations for bulk fused silica \cite{StaSaul98,NumTsu04}.

The sub-hertz OMIT feature is beneficial for various applications ranging from frequency-agile ultra-narrow filter 
cavities \cite{omit_gw_filter}, to coherent frequency converters \cite{TiaWan10,HillPain12},
to slow-light optical buffers \cite{Tuck05} and quantum memories. 
In the following we discuss the potential
of our system as a highly efficient slow-light buffer capable of seconds-long delays.
An optical buffer for coherent classical signals is characterized by the maximum possible delay that it 
can provide, and the usable bandwidth; they are not independent for passive systems, and in fact
the delay-bandwidth product (DBP) is bounded \cite{Tuck05,Khur05,Miller07,Bab08,Thev08}.
Further, if the buffer also features a near-unity storage and
retrieval efficiency, it may be used to store weak incoherent classical signals. In the limit that the signal
is encoded in a pure quantum state -- as required for various quantum information processing tasks -- 
the optical buffer becomes a quantum memory \cite{LvoTit09,AfzRei15}, if in addition to the above requirements, 
it also features a coherence time longer than the storage time.

\begin{figure}[t!]
\includegraphics[width=\columnwidth]{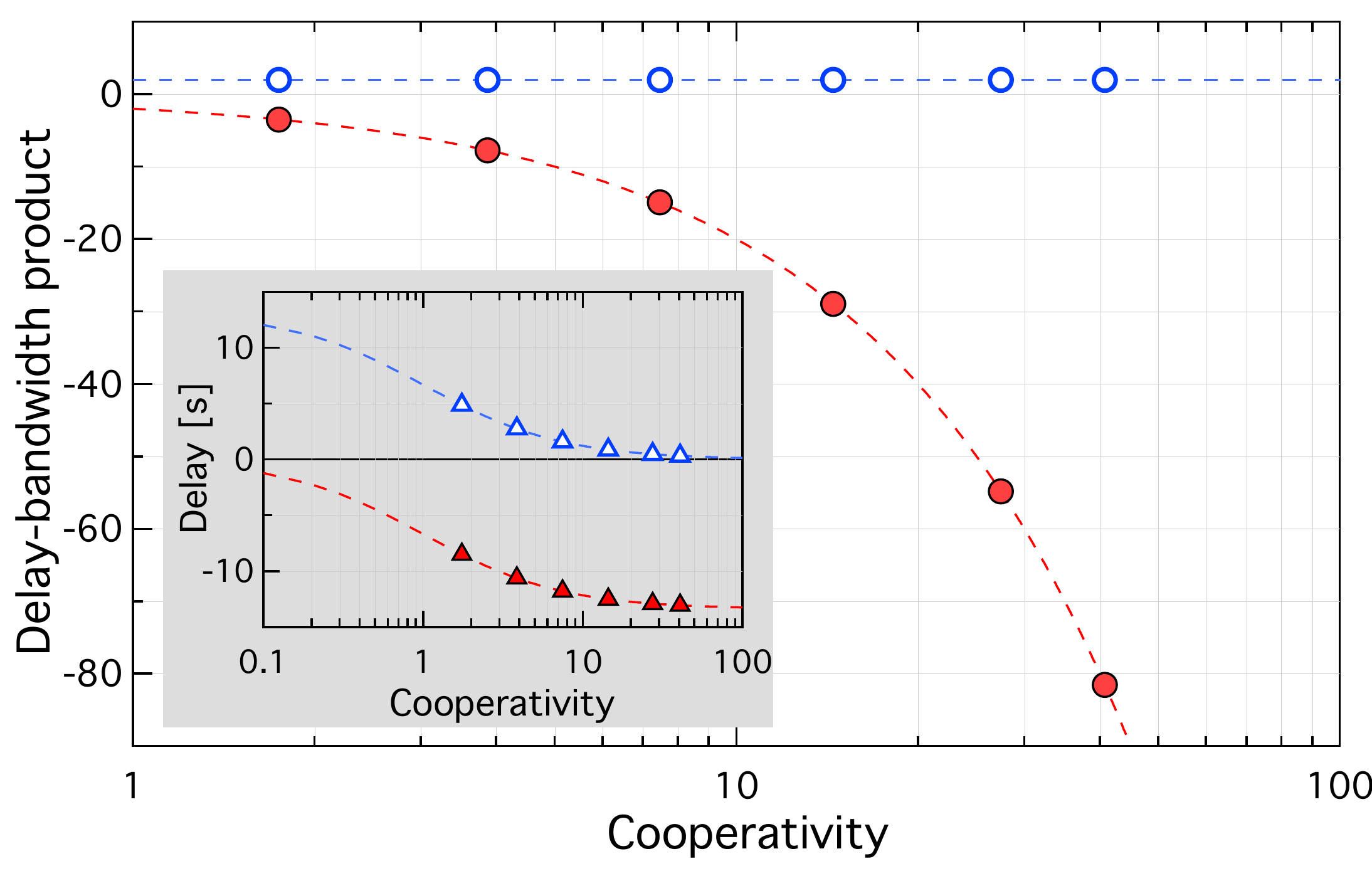}
\caption{\label{fig:dbp}
  Plot shows delay-bandwidth product in transmission (red) and reflection (blue); dashed lines show
  models based on \cref{eq:dbp}. 
  Inset shows the delay in transmission (red) and reflection (blue), with dashed lines showing models
  based on \cref{eq:tau}.
  }
\end{figure}

In the case of OMIT, the group delay, $\tau = [- \partial \phi[\Omega]/\partial \Omega]_{\Omega=\Omega_m}$, 
where $\phi$ is the phase response, is explicitly given by,
\begin{equation}\label{eq:tau}
  \tau = \frac{2C_\t{eff}}{\Gamma_\t{OMIT}}\times
    \begin{cases}
      -1 & \t{transmission} \\
      \frac{\eta_R}{1-\eta_R+C_\t{eff}} & \t{reflection}.
    \end{cases}
\end{equation}
It is negative (advance) 
or positive (delay) depending on whether the signal sideband is transmitted or 
reflected \cite{omit10,omit11,KarVit12}.
Note that $\eta_R = \kappa_\t{I}/\kappa \approx 0.999$ is the efficiency of the reflection port, given by
the fractional contribution of the input mirror decay rate $\kappa_\t{I}$ to the total cavity decay rate.
From the measured phase response, we are able to extract the delay in transmission and reflection, shown
in the inset of \cref{fig:dbp}. The inferred absolute delays, in the range of several (tens of) seconds, 
are more than an order of magnitude larger than what has previously been demonstrated using an optomechanical 
system \cite{KarVit12}, and approaching what has been demonstrated using atomic EIT \cite{HeiHalf13}.
Further, the signal efficiency in our system is near-ideal ($\eta_R \approx 0.999$), largely due
to the pristine optical quality, and significantly exceeds prior demonstrations of OMIT and even 
atomic EIT \cite{SchHalf16}. 
Finally, the combination of long delays, and near-ideal coupling efficiency, allows for a delay-bandwidth 
product (DBP) that is very large. From \cref{eq:tau}, the DBP is given by,
\begin{equation}\label{eq:dbp}
  \t{DBP} = \tau\cdot \Gamma_\t{OMIT} =
  2C_\t{eff}
  \begin{cases}
      -1 & \t{transmission} \\
      \frac{\eta_R}{1-\eta_R+C_\t{eff}} & \t{reflection}.
    \end{cases}
\end{equation}
The DBP takes a maximum value of $2$ for readout in reflection. \Cref{fig:dbp} shows that our system saturates 
this upper bound, which is also comparable with what is in principle achievable with atomic 
EIT systems \cite{Khur05,Tuck05}.
The potential of our system as a quantum memory is currently limited by the decoherence time of the mechanical 
mode ($\sim 0.2\, \t{\mu s}$), and the associated thermal noise. 
Recent work has demonstrated record low-noise quantum memory
using an intracavity Raman medium to suppress nonlinear mixing processes \cite{SauWalm16}, however at the
expense of efficiency.
In principle, optomechanical systems with macroscopic oscillators and low-loss mirrors can be free of
optical and mechanical nonlinearities -- respectively of the Duffing and thermal types -- while preserving
optical efficiency.
With further improvements employing
recently demonstrated techniques for mechanical Q enhancement \cite{TsaSchl17,GhadKip18}, it is conceivable that
the regime of quantum coherent mechanical oscillation ($Q \gtrsim n_\t{th}$, where $n_\t{th}$ is the
average thermal phonon occupation of the oscillator) can be achieved even in a
suspended-optic interferometer. In conjunction with adiabatically varying the 
pump amplitude \cite{FleLuk00,Tia12}, a long-lived on-demand OMIT-based quantum memory may be realized.

The combination of high cooperativity and ideal out-coupling efficiency are also 
the same requirements for using radiation pressure quantum fluctuations as a useful metrological resource
in an interferometer \cite{SudSch17}.
At the moment, the cooperativity we achieve is limited by a new source of angular instability.
Our observations are consistent with the conjecture that surface roughness on the cavity mirror leads to scattering
of the cavity's fundamental mode into a few higher order modes, which causes radiation pressure torques.
With mitigation of this problem we expect the current system to be a testbed for studying and reducing
the effects of quantum noise on a suspended interferometer with macroscopic test masses.

\emph{Acknowledgements.} This work was supported by the National Science Foundation via grants PHY-1707840 
and PHY-1404245. VS is supported by the Swiss National Science Foundation fellowship
grant P2ELP2\_178231.
We also gratefully acknowledge Eric Oelker and our colleagues in the LIGO Scientific Collaboration for valuable
interactions.
This paper has LIGO Document Number LIGO-P1800358.

\bibliographystyle{apsrev4-1}
\bibliography{omit_bib}


\clearpage
\begin{center}
  \textbf{\large Supplementary Information}
\end{center}

\tableofcontents

\section{Theoretical model for OMIT}

In the following we recapitulate the model used to interpret the OMIT data presented in
Figure 2 of the main manuscript. The presentation largely follows the standard treatment adopted in the 
cavity optomechanics community \cite{omit10,omit11}.

The basic optomechanical Hamiltonian that describes the radiation pressure interaction between the cavity 
field ($a$), end-mirror displacement ($x$), and the driving laser field ($a_\t{in}$) is \cite{AspKipMar14}
\begin{equation}\label{eq:Hom}
\begin{split}
  H &= \hbar \left( \omega_c -G\, x\right) a^\dagger a \\
   &\quad + \left(\frac{p^2}{2m} + \frac{m\Omega_m^2 x^2}{2} \right) \\
   &\quad + i \hbar \sqrt{\kappa_\t{I}} \left( a_\t{in}(t) a^\dagger - a_\t{in}^*(t) a \right),
\end{split}
\end{equation}
where,\\

\begin{tabular}{|l|l|}
  \hline
  $\omega_c$  &  bare cavity resonance frequency \\
  \hline
  $G$     &  bare optomechanical coupling strength \\
        &  $G \equiv \partial \omega_c /\partial x = \omega_c/L_c$  \\
  \hline
  $x,p$   & displacement and momentum of the oscillator \\
  \hline
  $m$     & effective mass of oscillator \\
  \hline
  $\Omega_m$  & frequency of oscillator \\
  \hline
  $a_\t{in}$  & input laser field (in units of $\sqrt{\text{photons}}/s$) \\
  \hline
  $\kappa_\t{I,E}$  & cavity loss rate through IM/EM \\
          &  $\kappa_\t{I,E}=\frac{c}{4L_c}T_\t{I,E}$; $T_\t{I,E}$ -- IM/EM transmission \\
  \hline
\end{tabular}\\

The Heisenberg equations that follow from \cref{eq:Hom} are,
\begin{equation}\label{eq:EOM}
\begin{split}
  \frac{da}{dt} &= \left( i\omega_c - \frac{\kappa}{2} \right)a - i G x a + \sqrt{\kappa_\t{I}} a_\t{in}\\  
  \frac{d x}{dt} &= \frac{p}{m} \\
  \frac{d p}{dt} &=  -m \Omega_m^2 x - \Gamma_m p - \hbar G a^\dagger a,
\end{split}
\end{equation}
where $\Gamma_m$ is the mechanical decay rate, and $\kappa = \kappa_\t{E}+\kappa_\t{int}+\kappa_\t{I}$ is the 
total loss rate of the cavity including from the EM, any internal loss, and the IM. 
Note that since the mechanical oscillator is high-Q ($Q_m \approx 10^6$), and since we are only interested in
its response near resonance, we have adopted a velocity-damped model for its loss. Further, since we are
interested in the driven response of the system, optical and mechanical input noises are omitted.

\noindent The input field that drives the cavity ($a_\t{in}$) is derived
from a laser oscillating at $\omega_\ell$ with sidebands $\delta a_\t{in}$ imprinted on it; 
it is thus described by,
\begin{equation}\label{eq:ainAnsatz}
  a_\t{in}(t) = \left(\bar{a}_\t{in} + \delta a_\t{in}(t) \right)e^{-i\omega_\ell t}.
\end{equation}
When \cref{eq:ainAnsatz} is substituted into the equation of motion in \cref{eq:EOM}, we arrive at a set of coupled
nonlinear equations that describe the full radiation pressure optomechanical dynamics which features
a static optical bistability and static spring shifts of the mechanical oscillator.
In the regime where the carrier flux is fixed and much larger than the sideband, 
i.e. $\abs{\bar{a}_\t{in}} \gg \abs{\langle\delta a_\t{in} \rangle }$, these equations can be linearized about a given operating point.
These linearized equations, expressed in the frame rotating at the laser frequency $\omega_\ell$, take the form,
\begin{equation}\label{eq:EOMlinear}
\begin{split}
  \left(\frac{d}{dt} -i\bar{\Delta} +\frac{\kappa}{2} \right) \delta a 
    &= -iG\bar{a}\, \delta x + \sqrt{\kappa_\t{I}} \delta a_\t{in} \\
  m\left(\frac{d^2}{dt^2} + \Gamma_m \frac{d}{dt} + \Omega_m^2 \right) \delta x 
    &= -\hbar G \bar{a}\left(\delta a + \delta a^\dagger \right).
\end{split}
\end{equation}
Here, we have defined an effective detuning,
\begin{equation*}
  \bar{\Delta} = (\omega_\ell - \omega_c) - G\bar{x},
\end{equation*}
that contains the bare laser-cavity detuning (first term) and a term due to the static cavity frequency shift
from radiation pressure coupling; $\bar{x}$ is the static mirror displacement, while
\begin{equation}\label{eq:abar}
  \bar{a} = \frac{\sqrt{\kappa_\t{I}}\,\bar{a}_\t{in}}{-i\bar{\Delta}+\kappa/2},
\end{equation}
is the mean intracavity field amplitude; and $\delta x, \delta a$ are the fluctuations on top of these
mean values. (Note that we have omitted the phase of the intracavity field in \cref{eq:EOMlinear} with the 
understanding that it is a constant offset from the phase of the input laser for fixed detuning.)
It is convenient to express \cref{eq:abar} in terms of the incident power, 
$P_\t{in} = \hbar \omega_\ell \abs{\bar{a}_\t{in}}^2$, and the mean intracavity photon number,
$n_c = \abs{\bar{a}}^2$, as,
\begin{equation}
  n_c = \frac{4\eta_\t{I}}{\kappa}\frac{P_\t{in}/\hbar \omega_\ell}{1+(2\bar{\Delta}/\kappa)^2}, 
\end{equation}
where we have defined, $\eta_\t{I} = \kappa_\t{I}/\kappa$, the cavity coupling efficiency from the incident port. 
Henceforth we will redefine $\bar{\Delta}\mapsto \Delta$ for notational convenience; further, we take $\bar{a}$ 
to be real by absorbing its (frequency-independent) phase, $\tan^{-1}(2\Delta/\kappa)$, into the input, with the
understanding that such static phase shifts are irrelevant in our measurement.

The OMIT phenomena entails a modification of the cavity response via its radiation pressure interaction 
with the end mirror. In the experiment, we measure the magnitude and phase of this modified response at frequency
offsets $\Omega$ from the incident pump laser, using probe sidebands at these frequencies, described by,
\begin{equation}
  \delta a_\t{in}(t) =  \delta A_\t{in}^+ e^{-i\Omega t} + \delta A_\t{in}^- e^{i\Omega t},
\end{equation}
where $\delta A_\t{in}^\pm$ are the amplitudes of the upper and lower sideband, respectively; 
since the sidebands are imprinted by phase modulation,
\begin{equation}\label{eq:Amod}
  \delta A_\t{in}^- = -\delta A_\t{in}^+ = -(\delta A_\t{in}^+)^* = (\delta A_\t{in}^-)^*.
\end{equation}
Such a drive produces intracavity fields and oscillator displacements at the same frequency since
the equations of motion \cref{eq:EOMlinear} are linear.
In order to track these we introduce the ansatz,
\begin{equation}
\begin{split}
  \delta a &= \delta A^+ e^{-i\Omega t} + \delta A^- e^{i\Omega t}\\
  \delta x &= \delta X e^{-i\Omega t} + \delta X^* e^{i\Omega t},
\end{split}
\end{equation}
into \cref{eq:EOMlinear}, and separate out terms oscillating at the two sideband frequencies;
this gives the closed set of coupled equations:
\begin{equation}\label{eq:AX}
\begin{split}
    \chi_c^{-1}[\Omega+\Delta]\, \delta A^+ 
      &=  -iG \bar{a}\, \delta X + \sqrt{\kappa_\t{I}}\, \delta A_\t{in}^+\\
    \chi_c^{-1}[\Omega-\Delta]\, (\delta A^-)^* 
      &= +iG \bar{a}\, \delta X +  \sqrt{\kappa_\t{I}}\, (\delta A_\t{in}^-)^*\\
    2m\Omega_m\, \chi_m^{-1}[\Omega-\Omega_m]\, \delta X 
      &= -i\hbar G \bar{a}\left(\delta A^+ + (\delta A^-)^* \right)
\end{split}
\end{equation}
where we have defined the optical and mechanical susceptibilities,
\begin{equation}\label{eq:chi}
  \chi_c^{-1}[\Omega] = \frac{\kappa}{2}-i\Omega, \qquad
  \chi_m^{-1}[\Omega] = \frac{\Gamma_m}{2}-i\Omega.
\end{equation}
Note that in going to \cref{eq:AX}, we have approximated the mechanical susceptibility using a 
single-pole response,
\begin{equation*}
\begin{split}
    m(\Omega_m^2 -\Omega^2 -i\Omega \Gamma_m ) 
      &= m\left((\Omega_m-\Omega)(\Omega_m+\Omega)-i\Omega \Gamma_m \right) \\
      &\approx m\, (-2i\Omega_m) \left( \frac{\Gamma_m}{2} + i (\Omega_m-\Omega) \right),
\end{split}
\end{equation*}
which is effectively a rotating-wave approximation valid in the high-Q limit, $Q_m = \Omega_m/\Gamma_m \gg 1$.

Solving \cref{eq:AX} for the mechanical motion excited by the intracavity field,
\begin{equation}\label{eq:dX}
\begin{split}
  2m\Omega_m \chi_{m,\t{eff}}^{-1}[\Omega] \delta X = -i\hbar G \bar{a} \sqrt{\kappa_\t{I}}
    &\left( \chi_c[\Omega+\Delta] \delta A_\t{in}^+\right. \\
    & \left. + \chi_c[\Omega-\Delta] (\delta A_\t{in}^-)^* \right)
\end{split}
\end{equation}
where, the effective mechanical susceptibility,
\begin{equation}\label{eq:chieff}
  \chi_{m,\t{eff}}^{-1}[\Omega] = \chi_m^{-1}[\Omega-\Omega_m] + g^2 (\chi_c[\Omega+\Delta]-\chi_c[\Omega-\Delta]),
\end{equation}
describes the radiation pressure modification of the mechanical response, whose strength scales with the
optomechanical coupling rate, $g = G\bar{a}\sqrt{\hbar/2m\Omega_m}$. 
When optomechanical coupling is weak enough to not lead to normal-mode splitting (i.e. when $g \lesssim \kappa$), 
the effective mechanical susceptibility can be approximated in terms of a modified mechanical 
linewidth (optical damping) and resonance frequency (optical spring), viz.,
\begin{equation*}
  \chi_{m,\t{eff}}^{-1}[\Omega-(\Omega_m+\delta \Omega_m)] \equiv \frac{\Gamma_m + \delta \Gamma_m}{2} 
  -i [\Omega -(\Omega_m + \delta \Omega_m)],
\end{equation*}
where $\delta \Gamma_m$ and $\delta \Omega_m$ are identified by separating the real and imaginary parts of the 
second term in \cref{eq:chieff}. 
We are interested in these expressions for the case of red-sideband pumping, i.e. $\Delta = -\Omega_m$;
in this case,
\begin{equation}
\begin{split}
  \frac{\delta \Gamma_m}{\Gamma_m} &= C\cdot \frac{(4\Omega_m/\kappa)^2}{1+(4\Omega_m/\kappa)^2} \\
  \frac{\delta \Omega_m}{\Omega_m} &= \frac{2C}{Q_m}\cdot \frac{4\Omega_m/\kappa}{1+(4\Omega_m/\kappa)^2},
\end{split}
\end{equation}
where, 
\begin{equation}
  C \equiv \frac{4g^2}{\kappa \Gamma_m},
\end{equation}
is the optomechanical cooperativity, which quantifies the fractional effect of the radiation-pressure modification
of the mechanical susceptibility. Note that in our case, characterized by $Q_m > C > 1$, the optical spring is weak
for red-sideband pumping, and we henceforth neglect $\delta \Omega_m$. The oscillator response is thus described
by $\chi_{m,\t{eff}}[\Omega-\Omega_m]$ featuring a modified linewidth.

The effect of the modified oscillator is to scatter phase-modulation sidebands on the intracavity field,
which can then interfere with the sidebands already present from the modulation imprinted on the incident field.
The resulting intracavity field can be obtained by inserting \cref{eq:dX} back into \cref{eq:AX} and solving for 
$\delta A^\pm$, viz.,
\begin{equation}\label{eq:dAsol}
  \delta A^\pm = \sqrt{\kappa_\t{I}}\, \chi_c[\Omega \pm \Delta]\, \delta A_\t{in}^\pm \times 
  \begin{cases}
    K[\Omega-\Omega_m] \\
    K^*[\Omega-\Omega_m],
  \end{cases}
\end{equation}
where, 
\begin{equation}\label{eq:K}
  K[\Omega] \equiv \frac{\chi_m^{-1}[\Omega]}{\chi_{m,\t{eff}}^{-1}[\Omega]}
  \approx 1- \frac{\delta \Gamma_m}{(\Gamma_m + \delta \Gamma_m) -2i\Omega}.
\end{equation}
In writing these expression, we have used the fact that the input laser is phase-modulated.
Note that in the absence of optomechanical coupling (i.e., $g=0$), $\chi_{m,\t{eff}}=\chi_m$, and
\cref{eq:dAsol} simply describes the cavity response $\chi_c$ filtering the incident field. 
The effect of optomechanical coupling is captured by the factor $K[\Omega]$
-- featuring the shape of the modified mechanical susceptibility -- which is superimposed on top of the 
(relatively) slowly varying cavity response.
It is this superimposed feature that manifests as an optomechanically-induced transparency window.
Note that as the optomechanical coupling is increased by strong pumping, $\delta \Gamma_m \gg \Gamma_m$, 
we have on resonance, 
$K[\Omega -\Omega_m]\vert_{\Omega \rightarrow \Omega_m} \rightarrow 0$, leading 
to complete transparency.

The experimentally observed fields are the ones leaking out of the cavity, either in transmission, or reflection.
The transmitted and reflected fields are \cite{GardColl87},
\begin{equation}\label{eq:IO}
\begin{split}
    a_\t{out,T}(t) &= \sqrt{\kappa_\t{E}}\, a(t) \\
    a_\t{out,R}(t) &= a_\t{in} - \sqrt{\kappa_\t{I}}\, a(t),
\end{split}
\end{equation}
where, $a(t) = \abs{\bar{a}}+\delta a(t)$, is the intracavity field in the rotating frame of the input laser. 
In the following we focus on the transmitted field, with the understanding that the reflected field can be
computed similarly. Using \cref{eq:dAsol,eq:Amod,eq:abar} in \cref{eq:IO}, the transmitted field is,
\begin{equation}\label{eq:aoutT}
\begin{split}
    a_\t{out,T}(t) = \sqrt{\kappa_\t{E}\kappa_\t{I}} \bar{a}_\t{in} &\left( \chi_c[\Delta] + \right. \\
    &\; \left. \chi_c[\Omega+\Delta]K[\Omega-\Omega_m] e^{-i\Omega t} \beta -\right. \\
    &\; \left. \chi_c[\Omega-\Delta]K^*[\Omega-\Omega_m] e^{+i\Omega t} \beta \right),
\end{split}
\end{equation}
where, $\beta \equiv \delta A_\t{in}^+/\bar{a}_\t{in}$, is the modulation index.
The first term in parenthesis describes the portion of the input pump that is transmitted, while the second
and third terms describe the upper and lower sidebands respectively. Eq. (2) of the main text is just
the second term, describing the transmission of the upper sideband alone. However, due to the finite sideband
resolution of our system, a sizable fraction, $\abs{\chi_c[\Omega-\Delta]/\chi_c[\Omega+\Delta]}$, of 
the lower sideband is also transmitted.
In our experiment, where the pump is red-detuned ($\Delta =-\Omega_m$) and at Fourier frequencies close to
mechanical resonance ($\Omega = \Omega_m$), this ratio is, $(\kappa/4\Omega_m)^2 \approx 3.8\%$.

When the transmitted field is detected on a photodetector, the detector output voltage is
$V_\t{T}(t) \propto \abs{a_\t{out,T}(t)}^2$; in the experiment we detect the voltage that is phase-coherent
with the input modulation. We are thus interested in the in-phase and quadrature-phase components of the
voltage oscillating at $\Omega$. This oscillating component is,
\begin{equation}\label{eq:dVout}
\begin{split}
    \delta V_\t{out,T}(t) &\propto \sqrt{\kappa_\t{E}\kappa_\t{I}}\,\t{Re}
    \left[ \chi_c[\Omega+\Delta]K[\Omega-\Omega_m] e^{-i\Omega t} \right. \\
    & \qquad\,\, \left. - \chi_c[\Omega-\Delta]K^*[\Omega-\Omega_m] e^{+i\Omega t} \right] \\
    & \equiv \delta V_\t{out,T}^\t{I}[\Omega] \cos (\Omega t) + \delta V_\t{out,T}^\t{Q}[\Omega] \sin (\Omega t).
\end{split}
\end{equation}
Note that here we have omitted the phase of $\chi_c[\Delta]$ which is a frequency-independent phase offset.
The final line in \cref{eq:dVout} implicitly separates out the in-phase and quadrature-phase components of the photodetector
signal measured by the network analyzer; the complex response -- the measured cavity transmission coefficient -- 
is then $ T[\Omega] \equiv \delta V_\t{out,T}^\t{I}[\Omega] + i\cdot \delta V_\t{out,T}^\t{Q}[\Omega]$.
Explicitly computing this gives,
\begin{equation}\label{eq:T}
  T[\Omega] = \sqrt{\kappa_\t{E}\kappa_\t{I}}\left( \chi_c[\Omega+\Delta]- 
    \chi_c^* [\Omega-\Delta]\right) K[\Omega-\Omega_m],
 \end{equation} 
which can be understood as the upper sideband contribution diminished by the undesired lower sideband transmitted
by the cavity. This is the full model used to fit the data in the main text. The reflection coefficient $R[\Omega]$
can be calculated in a similar fashion.

\subsection{Delay, bandwidth, and their product}

Both the cavity outputs, transmission and reflection, feature the conventional cavity response $\chi_c$ 
on top of which is superimposed the OMIT feature described by $K[\Omega]$. 
The former varies in frequency over a scale given by the cavity FWHM $\kappa$, while the latter varies 
within a much smaller interval, which can be read off from \cref{eq:K} to be the modified mechanical linewidth
$\Gamma_m + \delta \Gamma_m$. Thus the bandwidth of the OMIT feature is,
\begin{equation}
  \Gamma_\t{OMIT} = \Gamma_m + \delta \Gamma_m 
  \approx \Gamma_m \left(1+ C_\t{eff} \right),
\end{equation}
where we have defined the effective cooperativity,
\begin{equation}
  C_\t{eff} \equiv C\cdot \frac{(4\Omega_m/\kappa)^2}{1+(4\Omega_m/\kappa)^2},
\end{equation}
that characterizes the efficacy of dynamical radiation pressure effects for a system with finite sideband
resolution. In the limit of infinite sideband resolution, i.e. $\Omega_m \gg \kappa$, 
we have $C_\t{eff}\rightarrow C$, and the expression for $\Gamma_\t{OMIT}$ reduces to the one in the
literature \cite{omit10,omit11,Teuf11,Zhou13}.

The delay experienced by the probe depends on whether it is detected in transmission or reflection.
When detected in transmission, the resonant group delay is given by,
\begin{equation*}
  \tau_\t{T} = \left[-\frac{\partial}{\partial \Omega} (\t{arg}\, T[\Omega])\right]_{\Omega=\Omega_m}.
\end{equation*}
Evaluating this, using the expression for the transmission coefficient (\cref{eq:T}) and the expression 
for $K$ (\cref{eq:K}), we get,
\begin{equation}
\begin{split}
    \tau_\t{T} &= -\left( \frac{2}{\kappa} + \frac{2C_\t{eff}}{\Gamma_\t{OMIT}} \right) 
      \approx -\frac{2C_\t{eff}}{\Gamma_\t{OMIT}},
\end{split}
\end{equation}
where the first term is the delay due to the effect of the bare cavity, while the second term is
from OMIT. Since $\kappa \gg \Gamma_\t{OMIT}$ (in the weak-coupling regime we operate in), we 
can safely neglect the first term.

When detected in reflection, as in the case of single-port cavities \cite{omit10,omit11,Teuf11,Zhou13}, 
the group delay is affected by the out-coupling efficiency. To wit, 
\begin{equation}
\begin{split}
    \tau_\t{R} &= \left[-\frac{\partial}{\partial \Omega} (\t{arg}\, R[\Omega])\right]_{\Omega=\Omega_m} \\
      &\approx \frac{2 C_\t{eff}}{\Gamma_\t{OMIT}}\cdot \frac{\eta_\t{I}}{1+C_\t{eff}-\eta_\t{I}}.
\end{split}
 \end{equation}

The delay-bandwidth product (DBP),
\begin{equation}
  \t{DBP} \equiv \tau\cdot \Gamma_\t{OMIT},
 \end{equation} 
is thus different when the probe is measured in transmission or reflection. In fact,
\begin{equation}
  \t{DBP}_\t{T} = -2C_\t{eff}, \qquad \t{DBP}_\t{R} = \frac{2C_\t{eff}\eta_\t{I}}{1+C_\t{eff}-\eta_\t{I}}.
\end{equation}
Note that when the out-coupling efficiency through the reflection port is ideal ($\eta_\t{I}\rightarrow 1$),
$\t{DBP_R} \rightarrow 2$.

\section{Experimental details}

\subsection{Calibration of OMIT response}\label{sec:calib}

\begin{figure}[h!]
  \includegraphics[width=0.75\columnwidth]{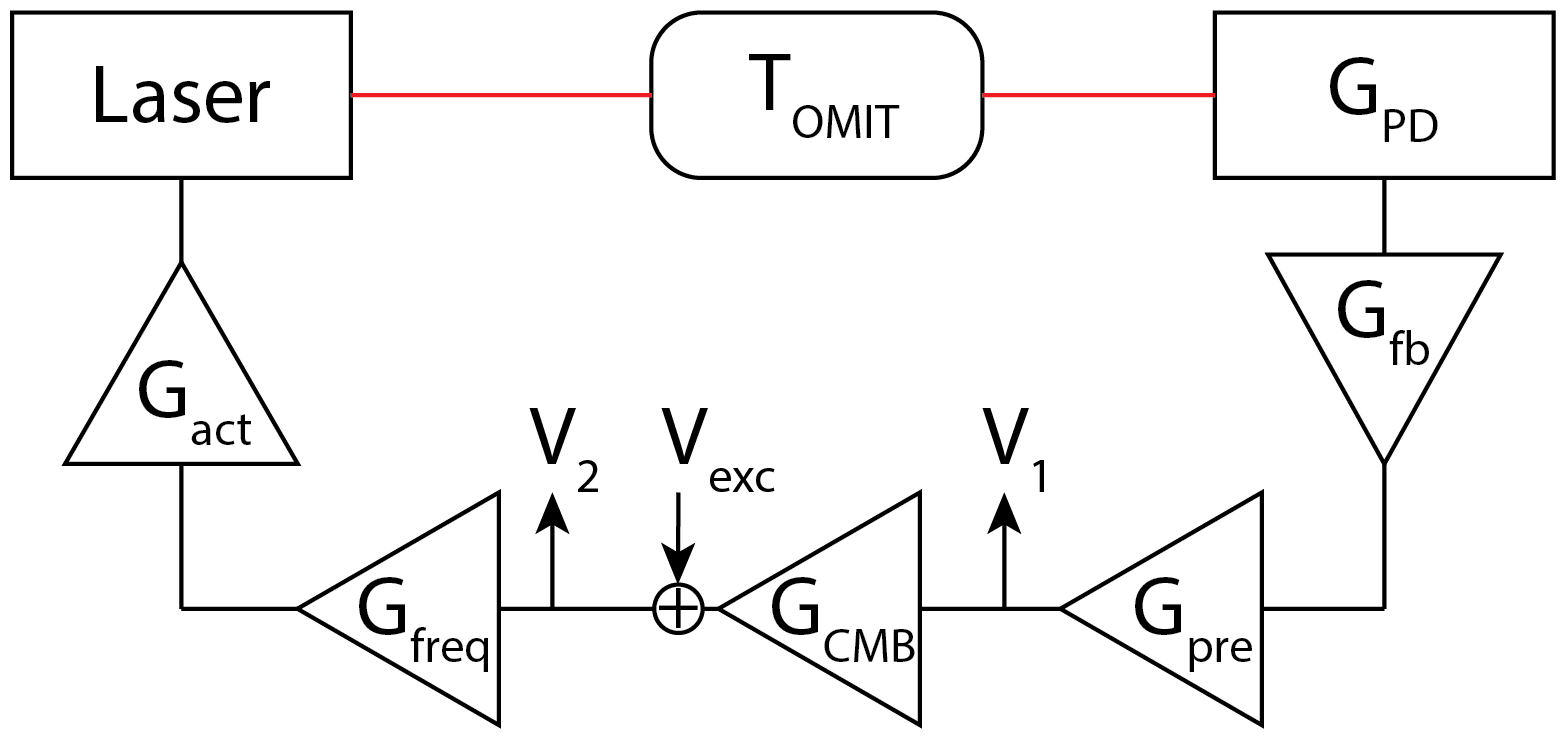}
  \caption{\label{fig:meas_loop_diagram}
    Loop diagram of the OMIT measurement. Red shows optical path, black shows electronic path.}
\end{figure}

In the experiment, both the excitation to probe the cavity and the readout, are done
inside the frequency stabilization servo loop, as shown in \Cref{fig:meas_loop_diagram}.
In order to isolate the OMIT response it is necessary to measure and calibrate out the responses of the
other elements of the loop. 

In the experiment, a network analyzer (SR785) is used to apply a stimulus $V_\t{exc}$ at the input of the
``common-mode board'' (CMB) -- a custom-built configurable electronic hardware \cite{SiggCMB} used as the length
control servo here --  which causes the laser frequency to modulate via the FSS (frequency stabilization servo) 
loop. This modulation is incident on the cavity, essentially sensing the OMIT response $T_\t{OMIT}$,
and gets detected at a photodiode (with response $G_\t{PD}$). The resulting signal is processed via an 
analog loop filter with response $G_\t{fb}$ and a pre-amplifier (SR560) with response $G_\t{pre}$.
A part of the output ($V_1$) is detected phase coherently with the excitation using the network analyzer,
while the rest is passed onto the CMB (with response $G_\t{CMB}$). A part of the CMB's output is also picked off
after being summed with the excitation ($V_2$) to be independently detected using the same network analyzer. 
This loop is shown in \Cref{fig:meas_loop_diagram}.

We use the network analyzer to measure the response $V_1/V_2$ (vis-a-vis the ratio of the two response
measurements $V_1/V_\t{exc}$ and $V_2/V_\t{exc}$).
From the loop diagram we can understand how to disentangle the information we need 
-- the OMIT response -- from this measurement.  
Going around the loop diagram we find for V$_1$:
\begin{equation*}
  (V_1 \cdot G_\t{CMB} + V_\t{exc}) 
  G_\t{freq} G_\t{act} T_\t{OMIT} G_\t{PD} G_\t{fb} G_\t{pre} = V_1,
\end{equation*}
which implies,
\begin{equation}\label{eq:omit_meas_v1}
  V_1 = V_\t{exc} \frac{G}{1-G} \frac{1}{G_\t{CMB}}.
\end{equation}
where, $G=G_\t{CMB}G_\t{freq}G_\t{act}T_\t{OMIT}G_\t{PD}G_\t{fb}G_\t{pre}$ is the open-loop
gain of the loop.
\begin{figure}[b!]
  \centering
  \includegraphics[width=\columnwidth]{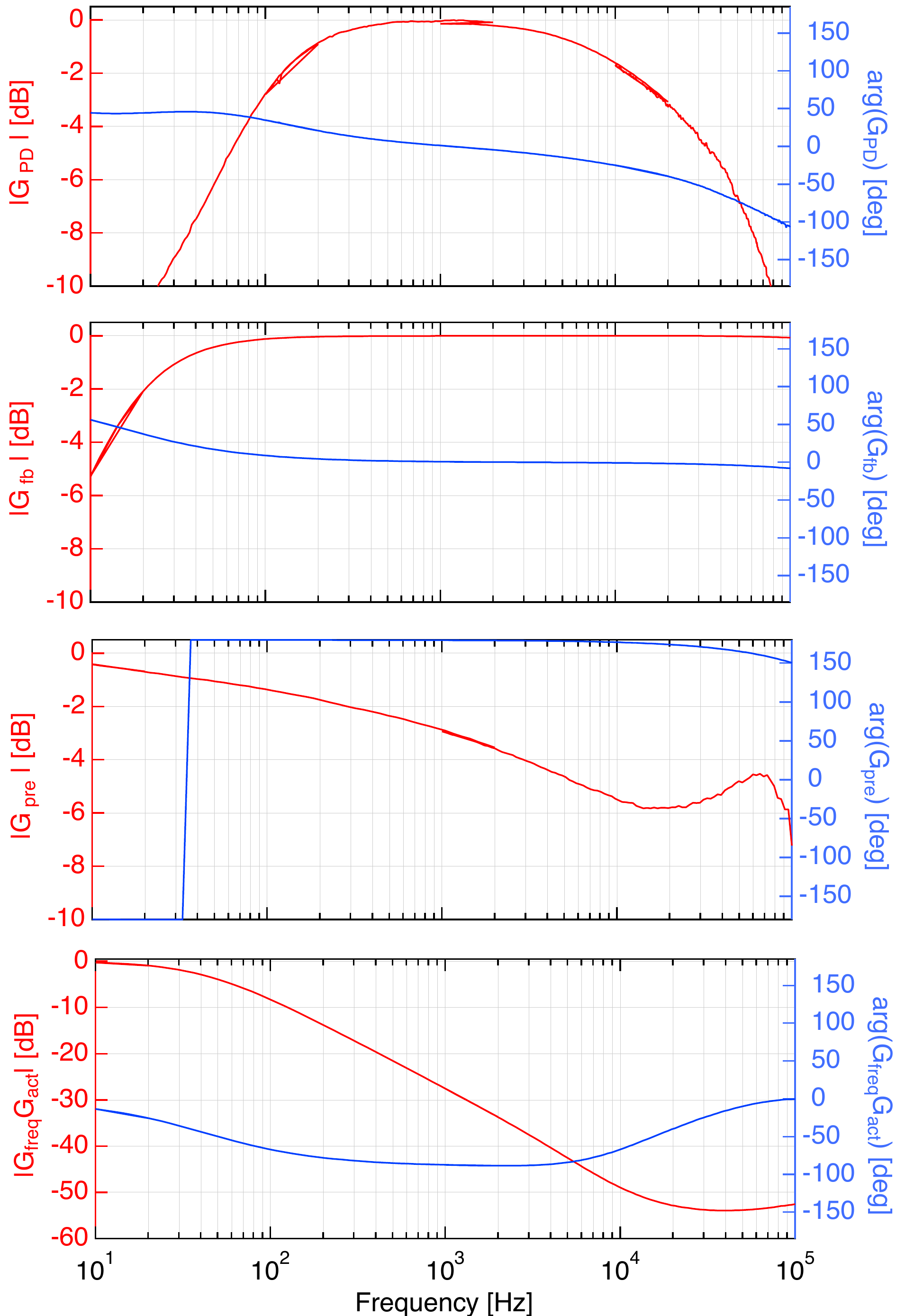}
  \caption{\label{fig:loop_responses}
    Measured responses of the various elements in the measurement loop. 
  }
\end{figure}
Performing a similar calculation for $V_2$ gives,
\begin{equation}\label{eq:omit_meas_v2}
\begin{split}
  V_2 = V_\t{exc} \frac{1}{1-G}.
\end{split}
\end{equation}
Combining \cref{eq:omit_meas_v1,eq:omit_meas_v2},
\begin{equation}
  \frac{V_1}{V_2} = \frac{G}{G_\t{CMB}},
\end{equation}
and so,
\begin{equation}\label{eq:corrected_omit_tf}
  T_\t{OMIT} = \frac{V_1/V_2}{G_\t{freq} G_\t{act} G_\t{PD} G_\t{fb} G_\t{pre}}
\end{equation}
\Cref{fig:loop_responses} shows the measured responses that are used in conjunction 
with \cref{eq:corrected_omit_tf} to infer the OMIT response.
Note that we are interested in the shape of the OMIT response, and not its absolute
magnitude which can be established from the independently measured broadband cavity transmission 
(shown in \cref{fig:cavity_response});
thus the measured responses in \cref{fig:loop_responses} omit overall dimensions for 
$G_\t{PD}$.

\begin{figure}[t!]
  \includegraphics[width=\columnwidth]{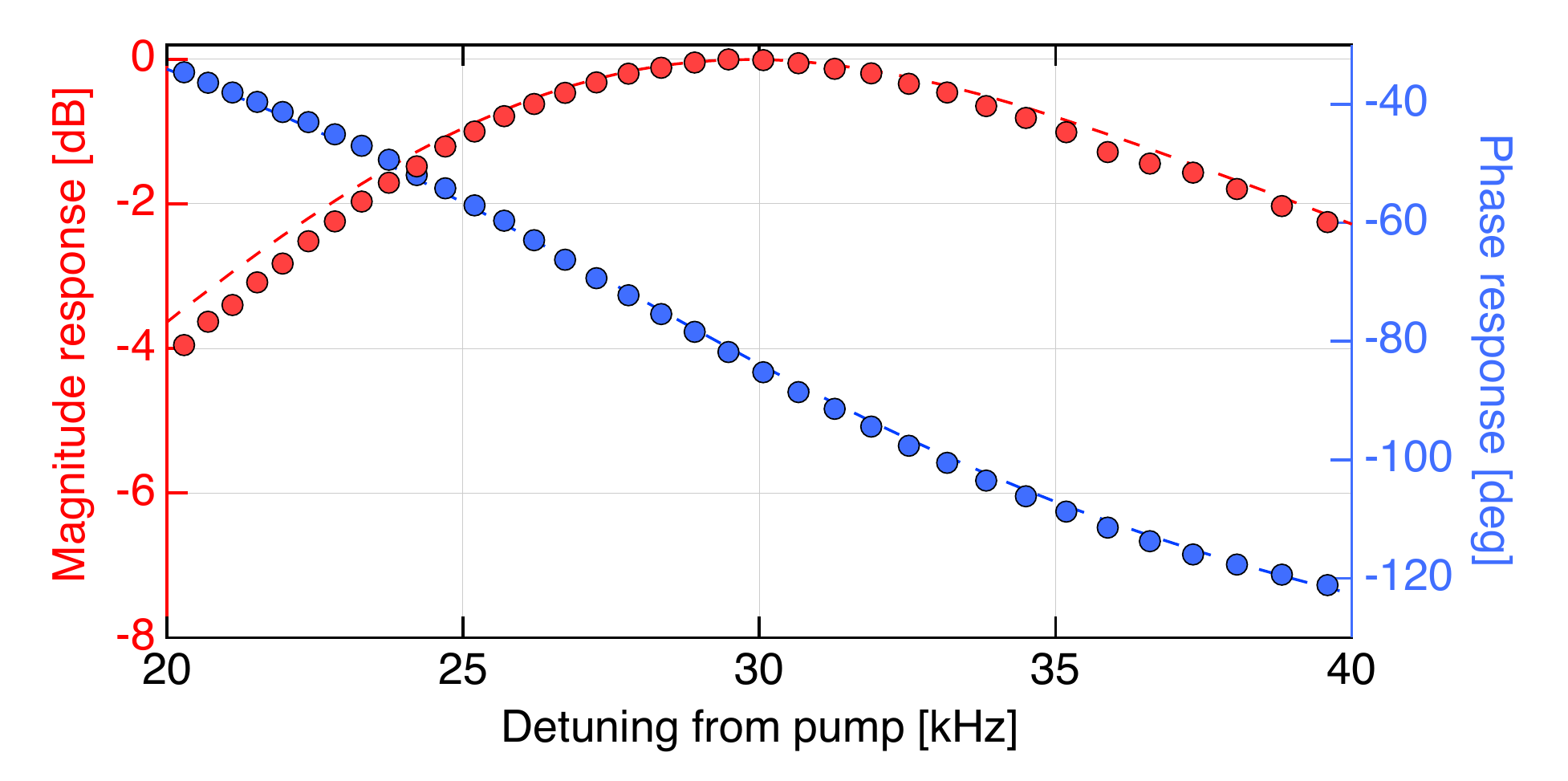}
  \caption{\label{fig:cavity_response}
    Example of a wide-band response measurement of the cavity taken at $60\, \t{mW}$ incident power.
    The dashed line shows fits. Note that due to the sparse sampling of frequency, the OMIT dip is 
    not visible in such measurements.}
\end{figure}

\subsection{Data analysis}

After the data has been corrected, we adopt the following procedure to extract the relevant optomechanical parameters.
The broadband cavity response, such as the one shown in \cref{fig:cavity_response}, is used to infer the detuning
$\Delta$ and the total cavity linewidth $\kappa$. 
Using these values, we then fit the narrowband cavity response,
containing the OMIT feature, to extract the mechanical frequency $\Omega_m$, its effective 
linewidth $\Gamma_m + \delta \Gamma_m$, and effective mass $m$, while the bare optomechanical coupling 
$G=\omega_c/L \approx 2\pi \cdot 0.28\, \t{GHz/\mu m}$ is assumed.
The fits to the data at each value of the incident power gives the following estimates for the various
parameters:
\begin{center}
\begin{tabular}{| l | l |}
  \hline      
  $\kappa/2\pi$ & $21.4 \pm 0.3$ kHz \\
  \hline
  $\Gamma_m/2\pi$ & $23.8 \pm 3.2$ mHz \\
  \hline 
  $\Omega_m/2\pi$ & $27.5$  kHz \\
  \hline
  $m$ & $133.7 \pm 9.6$ gm \\
  \hline 
  $\Delta/\Omega_m$ & $-0.96 \pm 0.01$ \\
  \hline  
\end{tabular}
\end{center}

\end{document}